\documentclass[onecolumn]{aa}
\usepackage{graphicx}
\usepackage{txfonts}
\usepackage{natbib}
\usepackage{placeins}
\newcommand{\solm}{M$_{\odot}$}
\newcommand{\rf}{\par\noindent\hangindent 15pt {}}

\begin{document}
%
\title{Polarimetry of Near-Infrared Flares from Sgr~A*}


   \author{A. Eckart\inst{1}
	  \and
          ~R. Sch\"odel\inst{1}
          \and 
          L. Meyer\inst{1}
          \and
          S. Trippe\inst{2}
          \and
          T. Ott\inst{2}
          \and
          R. Genzel\inst{2,3}
          }

   \offprints{A. Eckart}

   \institute{ I.Physikalisches Institut, Universit\"at zu K\"oln,
              Z\"ulpicher Str.77, 50937 K\"oln, Germany\\
              \email{eckart@ph1.uni-koeln.de}
         \and
             Max Planck Institut f\"ur extraterrestrische Physik,
              Giessenbachstra{\ss}e, 85748 Garching, Germany
         \and
             Physics Department,
             University of California at Berkeley, Le Conte Hall, 
             Berkeley, CA~94720, USA
             }

   \date{Received ; Accepted }

   \date{}

\abstract{
We report new polarization measurements of the variable near-infrared emission 
of the SgrA* counterpart associated with the
massive 3--4$\times$10$^6$\solm ~Black Hole at the Galactic Center.  
}{
We investigate the physical processes 
responsible for the variable emission from SgrA*.
}{
The observations have been carried out using the NACO adaptive
optics (AO) instrument at the European Southern Observatory's Very Large
Telescope\footnote{Based on observations at the Very Large Telescope
(VLT) of the European Southern Observatory (ESO) on Paranal in Chile;
Program: 271.B-5019(A).}.
}{
We find that the variable NIR emission of SgrA* is 
highly polarized and consists of a contribution of a non- or weakly 
polarized main flare with highly polarized sub-flares.
The flare activity shows a 
possible quasi-periodicity of 20$\pm$3 minutes consistent 
with previous observations.
}{
The highly variable and polarized emission supports that the NIR emission 
is non-thermal.
The observations can be interpreted in a jet or temporary disk model.
In the disk model the quasi-periodic flux density variations can be explained 
due to spots on  relativistic orbits around the central MBH. 
Alternative explanations for the high central mass concentration involving boson or
fermion balls are increasingly unlikely.
}


   \titlerunning{Polarized NIR emission from SgrA*}
   \authorrunning{Eckart et al. et al.}  
   \maketitle
%

\section{Introduction}
\label{section:Introduction}

Stellar dynamics has revealed the presence of a  
(3.6$\pm$0.3)$\times$10$^6$\solm ~Massive Black Hole (MBH) at
the center of the Milky Way 
at the position of the compact radio source SgrA*
(Eisenhauer et al. 2005, Ghez et al. 2005, 
Eisenhauer et al. 2003,
Sch\"odel et al. 2003, Eckart et al. 2002, Sch\"odel et al. 2002,
for a recent review see Eckart, Sch\"odel \& Straubmeier 2005).
The distance to this closest MBH is only 7.6$\pm$0.3 kpc
(Eisenhauer et al. 2005).
The close temporal correlation between
rapid variability of the near-infrared (NIR) and X-ray emission 
(Eckart et al. 2004, Belanger et al. 2005,
Eckart et al. 2006, and Yusef-Zadeh et al. 2006)
suggests that the emission with 10$^{33-34}$~erg/s flares  
arises from a compact source within a few ten 
Schwarzschild radii (R$_S$) of the MBH
(Baganoff et al. 2001, Baganoff et al. 2003, Eckart et al. 2004,
Genzel et al. 2003, Ghez et al. 2004,
Eisenhauer et al. 2005 and references therein).
This points to a common physical origin of the phenomena 
(Eckart et al. 2004, Eckart et al. 2006)
and may be linked to the variability at radio 
through sub-millimeter wavelengths
(Herrnstein et al. 2004, Mauerhan et al. 2005 and references therein).

The polarization properties of SgrA* in the radio to sub-mm wavelength 
regime are quite complex.
Between 1.4 and 112~GHz there is no significant linear
polarization, with upper limits of 0.1\%-2\%
(Bower et al. 1999a, 1999b, 1999c, Bower et al. 2001)
including a detection of weak ($\sim$0.4\%) circular polarization 
(Bower et al. 1999b). 
At higher frequencies of 150 to 400~GHz
Aitken et al. (2000) reported the first detection of about 10\% of
linear polarization from Sgr~A* 
(see also Bower et al. 2003, Melia \& Falcke 2001,
Bower et al. 1999a, Marrone et al. 2006).
From the lack of polarization at shorter frequencies, 
the authors concluded that the
polarized flux density arises in a compact 
millimeter/sub-millimeter component of Sgr~A*.

NIR/X-ray observations combined with theory indicate that the 
MBH source SgrA* is strongly variable with a  
NIR/X-ray spectrum 
involving Synchrotron self-Compton models
(Eckart et al. 2004, Eckart et al. 2006) and
rapid cooling of transiently heated electrons,
Doppler beaming, and modulation due to special relativistic
aberrations associated with the relativistic orbital motion 
(Broderick \& Loeb 2005, Eisenhauer et al. 2005, 
Ghez et al. 2004, Gillessen 2006)
or a short jet 
(Melia \& Falcke 2001, Markoff et al. 2001,
Falcke, Melia \& Agol 2000, Yuan, Markoff \& Falcke 2002).
Models show that polarized emission may arise 
in ordered toroidal magnetic fields of a 
disk-like accretion flow, where 
the E-vector of the 20\% to 60\% polarized radiation
may be perpendicular to the equatorial plane
(Broderick \& Loeb 2005,
Falcke, Melia \& Agol 2000,
Goldston, Quataert \& Igumenshchev 2005).
However, magnetic field configurations in accretion flows and (temporal) disks 
may have a strong turbulent component as well 
(Shakura \& Sunyaev 1973, Duschl, Balbus \& Hawley 1991, Balbus 2003).
At NIR wavelengths depolarization due to the 
Faraday effect or Thompson scattering can be neglected.
The NIR emission can then be expected to be significantly polarized, 
as already detected previously in a single weak flare (Genzel et al. 2003).

Here we report $\sim$3 minute time resolution NIR polarization measurements 
of a flare from the infrared counterpart of SgrA* which
clearly shows that the emission is non-thermal.
These data are of high interest because 
the polarized light in the NIR regime may carry information 
on the relativistic plasma close to the event horizon. 
During a $>$100 minute flare in July 2005 we detected 3 short sub-flares with
a polarization of up to 25\% and an associated swing in 
polarization angle of about $\sim$40$^o$
following the peak total and polarized intensities.

In section 2 we describe the observations and data reduction
and present the polarized NIR flux density light curves 
and polarization properties of SgrA* as a function of 
time in section 3.
The data is interpreted in the light of a possible jet and 
an orbiting spot model in sections 4 to 6.
A summary and conclusions are given in the final section 9.
A characterization of the observed variability as well as 
alternative models to the MBH hole scenario are discussed 
in the appendix.


\section{Observations and data reduction}
\label{section:Observations}

The observations of SgrA* have been carried out in the NIR K$_S$-band 
(2.0-2.36$\mu$m) using 
the  NIR camera CONICA and
the adaptive optics (AO) module NAOS on the 
European Southern Observatory's Very Large
Telescope UT4\footnote{Based on observations at the Very Large Telescope
(VLT) of the European Southern Observatory (ESO) on Paranal in Chile;
Program: 075.B-0093}
on Paranal, Chile, during the nights between 
12 and 13 June 2004 and 29 and 30 July 2005. 
The infrared wavefront sensor of NAOS was used to lock
the AO loop on the NIR bright (K-band magnitude $\sim$6.5) supergiant
IRS~7, located about $5.6''$ north of Sgr~A*.  
On both occasions the atmospheric conditions 
(and consequently the AO correction) 
were stable during the observations with optical seeing
values ranging between 0.5'' and 0.8''. 
Therefore the AO  could provide a stable correction with a high 
Strehl ratio (of the order 50\%).

In NACOS/CONICA (NACO) the combination of a Wollaston prism  
with a half-wave retarder plate 
allows the simultaneous measurement of two orthogonal 
directions of the electric field vector and the rapid change
between different angles, both crucial when trying to determine the 
polarization characteristics of a time varying source.
In July 2005 six and June 2004 four polarization angles 
could thus be observed in rapid sequence
within less than 3.5 minutes which is short 
compared to the typical time scales expected for flares
(Eckart et al. 2004, Genzel et al. 2003,
Ghez et al. 2004, Eckart et al. 2006).
Observations of a dark cloud 
a few arc-minutes to the north-west of Sgr~A* were interspersed with
exposures of the target in order to obtain sky measurements. 
All observations were dithered (in order to cover a
larger area and improve the image quality),
sky subtracted, flat-fielded, and corrected for
bad pixels. PSFs (point spread functions) were extracted from 
all the single exposures
using the {\it StarFinder} code
(Diolaiti et al. 2000).
The individual exposures were deconvolved with the Lucy-Richardson 
and with a Wiener filter algorithm and beam
restored with a Gaussian beam of FWHM corresponding to the respective
wavelength (60~milli-arcseconds at 2.2\,$\mu$m). 
Flux densities of compact sources were obtained via aperture photometry on the 
diffraction limited images.
These values were corrected for a 
background flux density contribution.
This background was determined as the mean value of fluxes
measured in the same size apertures at 5 different positions 
in a field $\sim$0.5'' to 1.5'' NW 
of SgrA* which is free of obvious contributions from individual stars.
Estimates of uncertainties were obtained from a comparison of these
results and from the standard deviation of fluxes of nearby constant sources
of similar brightness as SgrA*.
The calibration was performed
using polarization measurements of 
IRS21 ($\sim$13$\pm$4\% at PA$\sim$14$^o$) 
and the overall interstellar polarization of all sources in the field
of 3-4\% at $\sim$25$^o$ 
(Knacke \& Capps 1977, Lebofsky et al. 1982,
Ott et al. 1999, Eckart et al. 1995).

\section{Polarized NIR light curves from SgrA*}
\label{section:lightcurves}

In Figs.\ref{Fig:1} and \ref{Fig:2}
we show the 
flux density at different position angles (PA) 
of the Wollaston prism for a polarization reference star and
the NIR counterpart of SgrA*
in July 2005 and June 2004 with 6 and 4 
polarization angles, respectively.
These data clearly show that the polarization is significant
in certain parts of the overall light curve.
Since the overall light curve is termed 'flare', 
we give the shorter polarized events the term 'sub-flare' in the following.
Fig.\ref{Fig:3} (top right) shows the mean of our K-band NACO images 
of the central 0.6$\times$0.6 arcsec$^2$ and 
at 0$^o$ and 90$^o$ polarization angle 
taken simultaneously through the Wollaston prism 
during the peak of the flare at $t=$50 as plotted in Fig.\ref{Fig:1}. 
The polarization reference star (Fig.\ref{Fig:3}) is labeled as 'reference'.
The bottom right panel in Fig.\ref{Fig:3} shows the
difference between these 90$^o$ and 0$^o$ images clearly showing 
the excess of polarized de-reddened flux at the position of SgrA*
for the strongest sub-flare-1 in Fig.\ref{Fig:1}.
The top panels of Figs.\ref{Fig:4} and \ref{Fig:5} show
the total de-reddened flux density light curve and
the de-reddened flux density at polarization angle 90$^o$
as a function of time.
The bottom panels of these figures show the position angle 
of the E-vector and the degree of polarization.

In July 2005 (Fig.\ref{Fig:4}) the 
polarized sub-flares with FWHM durations of the order 10~min
are superimposed on top of the broader main flare.
The three sub-flares show peak polarizations of 12-25\%
with a swing in polarization angle of up to 40$^o$.
The most convincing swing occurs between $t=$50 and $t=$60 
in the July 2005 data.
At $t=$35 and $t=$75 the polarization is close to the galactic background 
polarization level of $\sim$3\% at $\sim$25$^o$ 
measured for the stars.
The most obvious sub-flares are those 
labeled sub-flares-1 and 
sub-flares-3 in Fig.\ref{Fig:4}.
Sub-flares-2 is weaker. 
It is a 2-3$\sigma$ event at position angles 
of 60$^o$ and 90$^o$ and a 1$\sigma$ event at a PA of 0$^o$.
At $t=$35-40 sub-flares-2 is followed by a local minimum at all PAs 
except 30$^o$. We therefore consider it as a significant sub-flare.

Marginal flux density excursions on the 1-2$\sigma$ level occur
at $t=$65, 75 and 85. 
The 2-3$\sigma$ flux density excursions at $t$$\sim$65 and $t$$\sim$75 
only occur at the orthogonal PAs of 60$^o$ and 150$^o$.
They therefore result from a single Wollaston prism setting
and are most likely due to a residual uncertainty in the flux 
density calibration of the individual channels.
We consider these excursions as not significant.

In general the polarized sub-flares are not clearly visible at polarization 
channels around $\sim$0$^o$ position angle which 
shows the least polarized flux density.
This indicates that the underlying broader flare is not significantly 
polarized (Figs.\ref{Fig:1} and \ref{Fig:4})
and that the sub-flares may be intrinsically 
highly polarized - possibly up to 60\%  
(Broderick \& Loeb 2005, Falcke, Melia \& Agol 2000,
Goldston, Quataert \& Igumenshchev 2005).

In June 2004 (Fig.\ref{Fig:2} and \ref{Fig:5}) a single 
polarized sub-flare with FWHM duration of the order 10~min
is superimposed on top of the broader main flare with a 
FWHM duration of about 20 minutes.
The first section of the overall flare ($t$=30-45) indicates 
that the broader underlying flare shows little ($\le$10\%) 
or no polarization. 
Its start preceeds the start of the 
significantly polarized
sub-flare at $t$$\sim$45 by about 10-15 minutes.
This is best seen in Fig.\ref{Fig:2}.
In Fig. \ref{Fig:5} (top) we compare the polarized flux density at
a PA of 90$^o$ to the total flux density of the flare.
In this figure the fact that the polarized sub-flare starts 
after the overall flare emission is less obvious since 
the total flux density contains both the polarized and unpolarized 
flux density.
With about 20\% the overall degree of polarization 
and a swing in position angle of about 40$^o$-50$^o$
the properties of the June 2004 sub-flare are very
much comparable to the sub-flares observed in July 2005.

Theoretical calculations indicate that 
the observed polarization properties of SgrA* 
may trace the relativistic accretion flow in a strong gravitational field
and that the polarization can be  used to probe the spacetime
at the photon orbit 
(Falcke, Melia \& Agol 2000,
Broderick \& Loeb 2005,
Goldston, Quataert \& Igumenshchev 2005).

\section{The jet model}
\label{section:jet}

Although in the case of SgrA* no jet has been positively detected
yet, our new NIR data could be taken as an indication for a
relativistic jet structure. The rotation of the polarization 
angle could be due to a helical magnetic field along the jet and the
variable sub-flare emission could be explained by temporal instabilities in the jet.

In compact extragalactic radio sources the distribution between the
E-vector position angles of the cores or jet components and the
jet axis show a rather flat distribution with the tendency 
of perpendicular orientations for QSOs
(Pollack, Taylor \& Zavala 2003) and a weak indication for stronger 
beamed sources to give a parallel arrangement
(Rusk 1988, Gabuzda, Pushkarev \& Cawthorne 2000, 
 but see also Pollack, Taylor \& Zavala 2003).

This can be understood in the framework of polarized radiation from
oblique and conical shocks (Cawthorne and Cobb 1990) and sources seen
at inclinations where the E-vector is parallel to the jet axis will
undergo the strongest variability due to relativistic beaming.
This is consistent with the finding of 
Gomez, Alberdi \& Marcaide (1994) who show that in helical jets, the strongest emission 
is obtained when the shock wave reaches regions where the jet is bent towards the observer. 
There the maximum Doppler boosting, the lengthening of the integration column, 
and the shock effects simultaneously contribute to the emission.
At these locations the electric vector lies parallel to the jet axis, 
due to the enhancement of the magnetic field parallel to the shock front. 
However, multi frequency observations of nuclear jets at cm-wavelengths also indicate 
polarization structure transverse to the jet axis (Pushkarev et al. 2005).
Along the bright spine of the jet the polarization E-vectors are aligned with the jet.
Within the `sheaths' at both edges of the jet the E-vectors are orthogonal 
to the jet axis.
The authors identify the presence of polarization aligned with the jet near the `spine' as an 
indication for the presence of a helical B-field that propagates outward with the jet flow. 

Polarized radio radiation and swings in position angles of several 10$^o$ have also
been observed for micro-quasars (Fender et al. 2000, 2003) and
low luminosity AGN (e.g. Bower et al. 2002ab).
The comparison to low luminosity galactic and luminous extragalactic sources shows
that a jet or outflow phenomenon possibly provides an explanation 
of our NIR data and that the E-vector may lie along the jet axis (or perpendicular to it).
A longer jet may have the potential of showing sub-flares that would stay
bright for a time exceeding the 10 minutes reported here.
Such a jet could be responsible especially for the emission 
at FIR/mm-wavelengths (Yuan, Markoff \& Falcke 2002).

There is not firm observational evidence for a jet or collimated 
outflow from SgrA* (see, however, discussion in section \ref{section:evidences}).
Therefore we have concentrated our interpretation on a disk model in
which the flare activity is due to luminous spots that are on
relativistic orbits around the central MBH.
In fact, near the last stable orbit (LSO)
a short jet emerging from a disk may likely look 
almost indistinguishable from a case involving a pure disk or orbiting spots.
This fact is supported by recent VLBI measurements of the intrinsic 
size of SgrA* at mm-wavelengths (see Fig.4 in Bower et al. 2004).
In addition the emission at the even shorter NIR wavelengths are likely to come
from even smaller scales in the jet model.

\section{The orbiting spot model}
\label{section:spot}

Our polarization measurements are consistent with
emission from 
highly polarized bright spots in relativistic 
orbits around the MBH 
(e.g.  Dovciak, Karas \& Yaqoob 2004,
Broderick \& Loeb 2005,
Goldston, Quataert \& Igumenshchev 2005).
We find that the sub-flares have a duration that is consistent with
the flux density amplification peaks expected for spots orbiting near the 
last stable orbit (LSO).
Therefore the sub-flares may arise from spots with a size 
of $\lesssim$R$_S$ embedded in a 
low polarization temporarily present (or bright) disk or ring
at a distance of a few R$_S$ possibly close to the LSO.
In such a two component model the disk would be responsible for the 
not significantly polarized overall flare and dilute the polarization 
signal from the higher polarized spots.
The time difference between the observed bright polarized sub-flares 
is about 20 minutes - similar to the quasi-periodicity oscillations (QPO)
reported previously (Genzel et al. 2003).
A total flux density periodogram of the July 2005 data shows
power at a period of 20$\pm$3 minutes (Fig.\ref{Fig:6}).
Given that from our data there is no evidence for red-noise excess power 
at periods well above 20 minutes beyond the 1-2$\sigma$ level
(see discussion on periodograms and red-noise in APPENDIX~1) 
we regard the power at a period of 20$\pm$3 minutes as significant. 
The stars show no excess power at a period of 20 minutes.
The top and bottom left panels in Fig.\ref{Fig:6} show the 
results of the analysis oversampled in order to allow a 
better estimate of the peak widths.

It is conceivable that the 3 sub-flare events observed in July 2005 
are associated with the same spot.
The required lifetime of a few 10 minutes to half an hour
can well be matched by the expected synchrotron cooling time scale
of a flare given via 
$t_s \sim 5 \times 10^8 \nu_9^{-0.5} B^{-3/2}$,
where $t_s$ is in seconds, B is in Gauss, $\nu_9$ is 
the self-absorption cutoff frequency of the synchrotron spectrum in GHz. 
For THz peaked spectra and typical magnetic fields
of several 10~G  
(Eckart et al. 2004, Eckart et al. 2006)
the synchrotron cooling time at the peak is of the order 
of 1 hour and represents an upper limit for the duration 
of the acceleration process that may also provide electrons 
radiating in the NIR.
In the NIR/MIR domain at frequencies around 
100~THz the cooling time scale would be of the order of the 
observed polarized sub-flare time scale. 
Under these circumstances an orbiting spot
could stay bright for a few orbital time scales 
(see Eckart et al. 2006 and Gillessen 2006).
The properties of the flare emission observed in June 2004 are
consistent with the above interpretation. The only difference is
that the overall flare event was about 5 times shorter and only one
polarized sub-flare was observed.

\section{On possible Evidence for Collimated Outflows}
\label{section:evidences}

The mean position angle of the E-vector may reveal information on the
orientation of the spin axis of the temporary disk or MBH
or even a possible jet or outflow.
Here we summarize the possible evidence that can be found for the 
jet and orbiting disk scenario.

\subsection{Evidence for a jet?}
\label{section:evidences-1}
In the case of SgrA* there is no direct observational evidence for a 
jet at radio and mm-wavelengths. 
However, the jet model is highly successful in explaining the radio 
observations (e.g. Bower et al. 2004).
There are, however, two conspicuous features that may be taken as an indication
for a highly collimated outflow approximately perpendicular to the position angle
of the E-vector. 
Morris et al. (2004) report on a thin elongated feature, 
a 'streak', in the CHANDRA X-ray image.
We labeled this feature XF in the inset of Fig.\ref{Fig:9}. 
This feature has a minimum length of 9'' (0.23~pc), is located at a mean 
distance of $\le$16.8'' ($\le$0.66~pc) to the SE (not SW as quoted by Morris et al. 2004) from
the center and points at a position angle of $\sim$30$^o$ to within $\le$5$^o$ into the direction of SgrA*.

In the thermal IR domain we find a comparable structure.
In addition to a few sharp edged structures associated with the
northern arm of the mini-spiral (e.g. Cl\'enet et al. 2004)
one can see an almost linear feature (LF) extending
over 2 thirds of an arcsecond and pointing 
at a position angle of $\sim$60$^o$ to within $\le$10$^o$ into
the direction of SgrA* (see Fig.\ref{Fig:9}). 
The NIR filament LF and the X-ray filament XF are not colinear.

The feature LF and an extended structure labeled EF in 
Fig.\ref{Fig:8} and \ref{Fig:9} are of special interest, 
as their geometry with respect to SgrA* rises the question
if they could be outflow signatures. 
LF indeed qualitatively shows similarity to a jet;
but the fact that it is bent and not pointing 
directly towards SgrA* may question this impression.
The EF is known as the "Finger" (Vollmer \& Duschl 2000)
or "Tip" of the Eastern arm to which it connects through a bubble-like
feature called the "microcavity" (Paumard et al. 2004).
The fact that EF is more difficult to identify in 
Fig.\ref{Fig:8} compared to Fig.\ref{Fig:9}
indicates that there is no filamentary structure contained
in EF that is as sharp as that of feature LF.
The dynamics of this feature were examined by Paumard et al. (2004),
who constructed velocity maps based on Br$\gamma$ line maps
obtained with integral field spectroscopy.
Those velocity maps show a high radial velocity ($\sim$300 km/s) of the
{\it Tip} and a very steep velocity gradient ($\sim$80 km/s) within this structure.
This clearly indicates, that EF is highly inclined with respect to the
plane of sky, leaving two possible interpretations:
\\
(1) The {\it Tip} is indeed an interaction zone between the Galactic Center ISM and
the top of a jet from SgrA*;
in this case it would be located behind the plane of the black hole.
\\
(2) The Eastern arm is a 'quiet' flow in front of the plane of SgrA*
which undergoes a dramatic loss of momentum due to the interaction with
a stellar wind at the location of the microcavity, and then starts
falling onto SgrA* through the {\it Tip}.
In that hypothesis, the microcavity could also be due to an
interaction not with a stellar wind, but with a jet. 
\\
In any case the {\it Tip} itself would then be the signature of an inflow 
rather than an outflow.
Combining the existing information, interpretation (2)
appears more probable than (1) as in Br$\gamma$ line maps
the northern edge of the Eastern arm appears more excited than
the rest of it. The cause for this is either a shock with the
Eastern bridge (which is definitely in front of the Northern arm,
if not of SgrA*), or because we mostly see the 'dark side' of the
Eastern arm, and this more excited edge pertains to the 'bright side'.

While the above described features observed in the X-ray and thermal IR
domain may indicate a collimated outflow 
such a phenomenon requires better proof.
An alternative explanation for these narrow and elongated features
are that they belong to the class of non-thermal filaments 
(NTFs; see e.g. LaRosa et al. 2004, Nord  et al. 2004 
and for reviews Morris 1996 Morris \& Serabyn 1996).

\subsection{The spin of a disk or MBH}
\label{section:evidences-2}

The mean position angle of the E-vector may be linked to
the orientation of the spin axis of the disk
that may be temporarily present (or bright) during the flares.
At small radii the flow's angular momentum may be 
tied to the MBH, which it accreted over its lifetime 
(Broderick \& Loeb 2005).
The flow orientation may also be set by the angular 
momentum of the stars whose winds feed the MBH
(Genzel et al. 2003, Levin \& Beloborodov 2003, see also 
Bardeen \& Petterson 1975).
For both data sets (July 2005 and June 2004) we 
observe a polarization PA that varies by $\pm$20$^o$
about a mean angle of $\sim$60$^o$ (E of N) and does not 
wrap around the 0$^o$/180$^o$ limit. 
To within the uncertainties this mean angle is the same for
all observed sub-flares.
If we assume that the sky projected magnetic field 
component parallel to the inclination axis of the disk 
is dominant then the distribution of PAs is not 
wrapped and the mean PA of the E-vector indicates 
the projected spin orientation on the sky.
Such a scenario may be present over most parts of an 
inclined disk with a toroidal field structure or in case of
a strong turbulent component with an isotropic distribution of 
magnetic field lines in or close to the disk plane
(Shakura \& Sunyaev 1973, Duschl, Balbus \& Hawley 1991, Balbus 2003).

Fig.\ref{Fig:7} shows model calculations of
the peak normalized flux density, degree of polarization, polarized 
flux density and PA of the E-vector as calculated for 
orbiting spot models 
based on the code by Dovciak, Karas, \& Yaqoob (2004).
The calculations were done for a single spot with a FWHM of about
one Schwarzschild radius, spin parameters of $a=$0.0, $a=$0.5, $a=$0.998, 
orbiting at the corresponding last stable orbits.
We assumed that the intrinsic E-vector is perpendicular (i.e. NS at a PA
 of 0$^o$) to the orbital plane. 
The calculations show that for 
inclinations $\gtrapprox$50$^o$ and spins below $a$$\lessapprox$0.5 the
polarization angle does not wrap around the 0$^o$/180$^o$ limit but
varies over the flare close to a value expected
for a non-relativistic orbiting disk 
(see also equation 4.6 in Bao et al. 1997). 

We find that in two independent data sets (July 2005 and June 2004) 
the PA does not wrap as expected
for inclinations $\gtrapprox$50$^o$ and spin values
$a$$\lessapprox$0.5. Under these conditions (see Fig.\ref{Fig:7}) 
the polarization angle varies over the flare close to 
a value expected for a non-relativistic orbiting disk 
and the mean orientations of the projected E-vector agrees
with that of the projected spin axis. 
This statement also holds for more sophisticated physical 
models of the intrinsic polarization
(Broderick \& Loeb 2005),
as long as the observed distribution of PAs is not
wrapped around the 0$^o$/180$^o$ limit.

Based on these theoretical predictions and on our observational data 
we therefore speculate that the measured E-vector of the polarized emission
may indicate the projected spin axis of a temporary disk around SgrA*.
In this picture an associated outflow would be perpendicular to the 
disk and therefore also perpendicular to the jet like features mentioned 
in section~\ref{section:evidences-1}.
Currently it is unclear how the direction of the projected E-vector 
can be related to the various structures observed in the central parsec.
In Fig.\ref{Fig:8} we show the range of orientations of the 
electric field vector on the sky. 
It points approximately into the direction of the
mini-cavity, which has been interpreted previously 
as being the result an interaction of the mini-spiral with 
a wind out of the general direction towards SgrA*
(Lutz et al. 1993, Melia et al. 1996, Yusef-Zadeh et al. 1998).
The projected E-vector is at large angles 
($\gtrapprox$50$^o$) with the spin vectors of all 
planar structures except the bar component and the system of 
clockwise orbiting young and luminous stars with high mass-loss 
(Levin \& Beloborodov 2003, Genzel et al. 2003,
Paumard et al. 2006 submitted to ApJ).
These planar structures may be related to a preferred plane
over which the SgrA* MBH accretes matter 
(Broderick \& Loeb 2005, Melia \& Falcke 2001,
Genzel et al. 2003, Levin \& Beloborodov 2003).

\section{A two component hot spot/disk model}
\label{section:model}

In order to demonstrate the consistency between the hot spot model 
with the observations we present a possible solution for the 
July 2005 measurements (A similar model can be obtained for the June 2004 data).
The QPOs find a natural explanation in a compact emission 
region that orbits around the MBH. 
In addition to the hot spot picture required to model the sub-flares,
we use a time varying underlying disk that 
generates the overall flare. 
For the demonstration purposes we have chosen to match 
the observed data on a visual basis. 
In what follows we describe how the modeling 
(see Fig.\ref{Fig:2}) was done and try to give constraints 
on the used parameters. 

The calculations have been carried out using the code by 
Dovciak, Karas \& Yaqoob (2004). An optically thick, geometrically 
thin emission region that lies within the equatorial 
plane has been assumed. 
For the intrinsic emissivity of the disk we chose a Gaussian 
shaped time dependency and a $r^{-1}$ spatial dependency
(following Yuan, Quataert \& Narayan 2004).
We assumed a polarization degree of 3\%. Higher intrinsic 
polarization would not allow for the observed changes 
in the polarization angle. 
The spot in our case has a radial extent of one $R_S$ 
and an orbital Gaussian shaped extent of 
$\sigma \sim 1.5 \, R_S$. 
The size was chosen in such a way that on the one hand 
the peaks in the flux curve are matched well and 
that on the other hand the changes in the polarization 
angle are not too small. The center of the spot is 
roughly located on the LSO so that for a MBH 
with $a=0.5$ the orbital period of $\sim 20$ min is matched. 
We have not considered shearing of the spot and assume that it
remains confined over a few orbits.
In order to match the variable observed flux density of the 
sub-flares, we changed 
the intrinsic luminosity for every period. 
Normalized to sub-flare-1 (Fig.\ref{Fig:4}), the spot has 50\% intensity 
during the first period (corresponding to sub-flare-3), 
20\% in the second (corresponding to sub-flare-2),
and 10\% in the fourth (corresponding to $t$$\sim$64). 
During the third revolution, the spot has a peak over-brightness 
relative to the disk (at the spot's location) of roughly 
50\% at the chosen inclination of $55^{\circ}$. 
We used this inclination, because in our model the polarization 
angle does not wrap around the $0^{\circ}/180^{\circ}$
limit only for viewing angles $\geq 50^{\circ}$ 
(see also Bao et al. 1997). 
Here we have assumed that the spot has an intrinsic linear 
polarization of 60\%
(Broderick \& Loeb 2005, Falcke, Melia \& Agol 2000,
Goldston, Quataert \& Igumenshchev 2005).
We chose the E-Vector to be perpendicular to the 
equatorial plane as might be expected for certain magnetic field
configurations (see comments above).
In addition to the intrinsic polarization
we include a 3\% foreground polarization along 30$^o$ (see references above).
The combined degree of polarization  
due the disk, spot and the foreground yield a degree of polarization of
up to 20\% and polarization angles that range between 
38$^o$ and 83$^o$, which matches what is inferred from the observations.  

As can be seen in Fig.\ref{Fig:10} our model represents 
the observed light curve reasonably well within the measurement uncertainties. 
Regarding the fact that we neglected possible tilts and warps of the
accretion disk (e.g. Bardeen \& Petterson 1975, Lubow, Ogilvie \& Pringle 2002),
magnetic field effects and used a highly idealized and 
therefore somewhat unphysical intrinsic polarization model, 
the detailed discrepancy of the polarization angle is not surprising.

\section{Summary and Conclusions}
\label{section:summary}

New high angular resolution NIR observations have shown that the flare
emission from SgrA* is highly polarized and provide a direct proof of the
non-thermal nature of the radiation.
At two observing epochs in July 2005 and June 2004 the mean position angle
of the E-vector was at about 60$^o$.
The data can be interpreted in a jet and orbiting disk model.
We have discussed possible features that may be indicative 
for a collimated out-flow or jet from SgrA*.
We also find new evidence for a possible quasi-periodicity of 20$\pm$3 minutes.
The new data strengthens the case for a MBH rather than for alternative explanations
of the high mass concentration (see discussion in APPENDIX~2).
In the model of compact ($\lesssim$R$_S$) highly polarized
hot spots orbiting SgrA* in a temporary disk
(see also Genzel et al. 2003)
at a radius of a few times the event horizon 
(Dovciak, Karas \& Yaqoob 2004,  Broderick \& Loeb 2005)
this period would be consistent with a MBH spin parameter of $a=0.5$.
This supports the presence of a Kerr MBH that rotates at about 
half its maximum spin and the position angle of the polarized emission
may indicate its projected orientation on the sky.


\vspace{0.2cm}
\noindent
{\bf Acknowledgments:}
This work was supported in part by the Deutsche Forschungsgemeinschaft
(DFG) via grant SFB 494.  
We are grateful to all members
of the NAOS/CONICA and the ESO PARANAL team,
especially N. Ageorges for support in setting up
the polarization measurements.  
We thank 
M. Dovciak,
A.E. Broderick, and A. Loeb
for very helpful discussions 
and especially 
M. Dovciak for valuable conversations and 
for providing the code to us.
K. Muzic provided the high pass filtered L'-band image.

\newpage

\begin{center}
\Large
{\bf APPENDIX}
\end{center}
\normalsize

\begin{center}
\large
{\bf APPENDIX 1: Characterization of the flux density variations}
\end{center}
\normalsize

The analysis of variability data by 
Mauerhan et al. (2005) for SgrA* at 3~mm wavelength 
and by (Benlloch et al. 2001) for X-ray data of Mrk~766 
have shown that in these cases the flux density
variations can well be described by a red-noise power spectrum of
the form $P(f) \sim f^{-\beta}$. 
In order to characterize the variations in 
our (densely and almost exclusively regularly sampled) 
polarized NIR flux density data
we calculated periodograms using the IDL LNP$\_$TEST function
that computes Lomb Normalized Periodograms and is based on the 
routine 'fasper' in section 13.8 of 
Numerical Recipes in C: The art of Scientific
Computing (Cambridge University Press).

To take into account the swing in polarization angle the periodograms 
were first obtained for data taken at the different individual
polarization angles and then averaged. 
The 1$\sigma$ error bars are derived as the standard deviation
of the results for the different polarization angles.
The excess power is better visible for angles with high polarized flux.
Therefore the 1$\sigma$ error bars are upper limits for the uncertainties
of the power values.
The result for SgrA* is shown as a black line in Fig.\ref{Fig:11}. 
In Fig.\ref{Fig:6} we also show in red the
mean and standard deviation calculated from periodograms of 7 stars 
within 1.5 arcsecond radius of SgrA*.

The SgrA* data can be fitted by a power spectrum with an exponent
$\beta = 1.2$ having a 1$\sigma$ uncertainty of 0.7.
Hence our July 2005 NIR data are consistent with a flat noise distribution
($\beta = 0$) and
do not provide an indication for a significant low frequency red-noise excess. 
The case for red-noise is weakened even further by the fact
that the power peaks at frequencies around 0.04 and 0.1 min$^{-1}$
(25 and 10 minute periods) 
can be attributed to the overall unpolarized part of the flare 
and a noise pattern that is also contained in the 
stellar periodograms, respectively.
The excess power around 0.1 min$^{-1}$ (10 minute period) 
corresponds to the frequency 
of dithering and therefore most likely 
derives from residual uncertainties
of the calibration and data processing.
Both SgrA* and the stars show a similarly shaped 
excess between frequencies 
of 0.2 to 0.07~min$^{-1}$ (5 to 15 minute periods)
and in Fig.\ref{Fig:11} the two periodograms have 
been normalized to the excess power in this range.
There is a  remaining relative scatter between the 
two curves of 1$\sigma$$\sim$0.25. 
The resulting 3$\sigma$ limit for intrinsic 
(combined polarized and unpolarized) power 
of SgrA* in this frequency range is shown 
as a bold face upper limit.

The excess power contained in the SgrA* data
at a frequency of 0.05~min$^{-1}$ (a 20 minute period) 
is not present in the stellar data.
We identify this excess with the power contained in the 
polarized portion of the SgrA* flare. 
In Fig.\ref{Fig:11} we plot in bold face the power contained in the 
light curve of SgrA* at this frequency corrected for the 
power contribution also present in the stellar light curves.
 
Using the fact that the relative powers relate as the integrals
under the corresponing portions of the light curves 
we can correct for the power contained in the unpolarized part 
of the overall flare emission at a frequency of 0.05~min$^{-1}$. 
We then obtain an upper limit for the polarized power at that
frequency which is shown in bold face in Fig.\ref{Fig:11}.
This number is an upper limit since the power contained
in the polarized portion of the light curve clearly occurs at
higher frequencies.

Since the SgrA* periodogram does not contain 
a dominant contribution of red-noise
the significance standards of Horne and Baliunas (1986)
apply and we regard the excess power at a frequency of 
0.05~min$^{-1}$ (20 minute period) as a significant indication for
a quasi-periodicty.
The physical interpretation of this excess is discussed 
in the paper sections
\ref{section:lightcurves}
and 
\ref{section:model}.

\begin{center}
\large
{\bf APPENDIX 2: Alternative models}
\end{center}
\normalsize

Explaining SgrA* with alternative solutions for a MBH becomes 
increasingly difficult. Stellar orbits near SgrA* make a 
universal Fermion ball solution for compact galactic nuclei
highly unlikely 
(Sch\"odel et al. 2002, Maoz 1998, see also Melia \& Falcke 2001
and Yuan, Narayan \& Rees 2004).
The good agreement between the measured polarized flare structure 
and the theoretical predictions 
(Broderick \& Loeb 2005, Eisenhauer et al. 2005,
Ghez et al. 2004, Gillessen 2006)
as well as the indication of a quasi-periodicity in the data 
severely challenge the alternative explanation of the central 
mass as a massive boson star
(Torres, Capozziello \& Lambiase 2000, Mielke \& Schunck 2000, 
Lu \& Torres 2003 and references therein).
If an {\it ad hoc} weak repulsive force between a hypothetical brand
of bosons is introduced, it appears to be possible to form massive,
compact objects with sizes of a few R$_S$ that
are supposedly supported by the Heisenberg uncertainty principle. 
However, it is a delicate process to form a boson star and 
preventing it from collapsing to a 
MBH despite of further accretion of matter, 
a non spherically symmetric arrangement of
forces as in the case of a jet or matter 
being in orbit around the center but well within the boson star.
Such a massive boson star scenario could already be excluded for the 
nucleus of MCG-6-30-15 (Lu \& Torres 2003).
In the case of a stationary boson star the orbital velocity 
close to the $\sim$3~R$_S$ radius LSO
is already $\sim$3 times lower than that of a 
Schwarzschild MBH (Lu \& Torres 2003)
and relativistic effects are severely diminished 
and further reduced at even smaller radii.
If the indicated quasi-periodicity is due to orbital motion 
then a stationary boson star can be excluded
as an alternative solution for SgrA*, since in this case 
one expects the orbital periods to be larger.

\newpage

%
%
%

\vspace*{1cm}

\rf{Aitken, D.K., Greaves, J., Chrysostomou, A., Jenness, T., Holland, W., 
  Hough, J.H., Pierce-Price, D., Richer, J., 2000, ApJ 534, L173}
\rf{Baganoff, F.K., Bautz, M.W., Brandt, W.N., et al. 2001, Nature, 413, 45}
\rf{Baganoff, F. K., Maeda, Y., Morris, M., et al. 2003, ApJ 591, 891}
\rf{Balbus, S.A., \& Hawley, F.H., 1991, ApJ 376, 214}
\rf{Balbus, S.A., 2003, ARA\&A 41, 555}
\rf{Bao et al., 1997, ApJ, 487, 142}
\rf{Bardeen, J.M. \& Petterson, J.A., 1975, ApJL 195, L65}
\rf{Belanger, G., Goldwurm, A., Melia, F., Yusef-Zadeh, F., Ferrando, P., Porquet, D., Grosso, N.,
  Warwick , R., 2005, ApJ 635, 1095, astro-ph/0508412}
\rf{Benlloch, S., Wilms, J., Edelson, R., Yaqoob, T., \& Straubert, R., 2001, ApJL 562, L124}
\rf{Bower, G.C., Wright, M.C.H., Backer, D.C., Falcke, H., 1999a, ApJ 527, 851}
\rf{Bower, G.C., Falcke, H., Backer, D.C., 1999b, ApJ 523, L29}
\rf{Bower, G.C., Backer, D.C., Zhao, J.-H., Goss, M., Falcke, H., 1999c, ApJ 521, 582}
\rf{Bower, G.C., Wright, M.C.H., Falcke, H., Backer, D.C., 2001, ApJ 555, L103}
\rf{Bower, G.C., Wright, M.C.H., Falcke, H., \& Backer, D.C. 2003, ApJ 588, 331}
\rf{Bower, G.C., Falcke, H., Mellon, R.R., 2002a, ApJL 578, L103}
\rf{Bower, G.C., Falcke, H., Sault, R.J., Backer, D.C., 2002b, ApJ 571, 843}
\rf{Bower, G.C., Falcke, H., Herrnstein, R.M., Zhao, J.-H., Goss, W. M., Backer, D.C., 2004, Science 304, p. 704.
\rf{Broderick, A.E. , Loeb, A., 2005, astro-ph/0509237 }
\rf{Cl\'enet, Y. et al., 2004, A\&A 417, L15}
\rf{Diolaiti, et al. 2000, A\&A Suppl. 147, 335}
\rf{Dovciak, M., Karas, V., \& Yaqoob, T. 2004, ApJS 153, 205}
\rf{Eckart, A., et al., 2006, A\&A in press; astro-ph/0512440}
\rf{Eckart, A., Sch\"odel, R., Straubmeier, C., 2005,
	'The black hole at the center of the Milky Way',
         London: Imperial College Press}
\rf{Eckart, A. et al., 2004, A\&A 427, 1}
\rf{Eckart, A., Genzel, R., Ott, T. and Sch\"odel, R. 2002, MNRAS, 331, 917}
\rf{Eckart et al. 1995 ApJL 445, L23}
\rf{Fender, R., Rayner, D., Norris, R., Sault, R.J., Pooley, G., 2000, ApJL 530, L29}
\rf{Fender, R., 2003, Astrophysics and Space Science 288, 79}
\rf{Eisenhauer, F.; Sch\"odel, R.; Genzel, R.; Ott, T.; Tecza, M.; Abuter, R.; Eckart, A.; Alexander, T., 2003, ApJ 597, L121}
\rf{Eisenhauer, F. et al.  2005, ApJ 628, 246}
\rf{Falcke, H., Melia, F., Agol, E., 2000, ApJL 528, L13}
\rf{Gabuzda, D.C., Pushkarev, A.B., Cawthorne, T.V., 2000, MNRAS 319, 1109}
\rf{Genzel, R., Sch\"odel, R., Ott, T., et al. 2003, Nature, 425, 934}
\rf{Genzel, R., et al., 2003, ApJ 594, 812}
\rf{Ghez, A.M., Wright, S.A., Matthews, K., et al. 2004, ApJL 601, L159}
\rf{Ghez, A.M., Salim, S., Hornstein, S. D., Tanner, A., Lu, J. R., Morris, M., Becklin, E. E., Duch\^ene, G.,2005, ApJ 620, 744}
\rf{Gillessen, S., et al. 2006, ApJ 640, L163}
\rf{Goldston, J.E.,  Quataert, E.,  Igumenshchev, I.V. 2005, ApJ 621, 785}
\rf{Gomez, J. L., Alberdi, A., Marcaide, J. M., 1994, A\&A 284, 51}
\rf{Herrnstein, R.M., Zhao, J.-H., Bower, G.C., \& Goss, W.M., 2004, AJ, 127, 3399}
\rf{Horne, J.H., \& Baliunas, S.L., 1986, ApJ 302, 757}
\rf{Knacke, R. F., Capps, R. W., 1977, ApJ 216, 271}
\rf{LaRosa, T.N., Nord, M.E., Lazio, T.J.W., Kassim, N.E., 2004, ApJ 607, 302-308}
\rf{Lebofsky, M.J., Rieke, G.H., Deshpande, M.R., Kemp, J. C., 1982, ApJ 263, 672}
\rf{Levin, Yuri; Beloborodov, Andrei M., 2003, ApJL 590, L33	}
\rf{Lu, Y., \& Torres, D.F., 2003, Int. Journal of Modern Physics D, Vol.12, No. 1, pp. 63-77}
\rf{Lubow, S.H., Ogilvie, G.I., Pringle, J.E., 2002, MNRAS 337, 706}
\rf{Lutz, D., Krabbe, A., Genzel, R, 1993, ApJ 418, 244}
\rf{Mauerhan, J.C.; Morris, M.; Walter, F.; Baganoff, F.K., 2005, ApJL 623, 25}
\rf{Markoff, S., Falcke, H., Yuan, F. \& Biermann, P.L.  2001, A\&A, 379, L13}
\rf{Maoz, E., 1998, ApJL 494, L181}
\rf{Marrone, D.P., Moran, J.M., Zhao, J.-H., Rao, R., 2006, ApJ in press, astro-ph/0511653}
\rf{Melia, F., Coker R.F., Yusef-Zadeh, F., 1996, ApJ 460, L33}
\rf{Melia, F., Falcke, H., 2001, ARA\&A, 39, 309}
\rf{Mielke, E., \& Schunck, F., 2000, Nucl. Phys. B 594, 1985}
\rf{Morris, M., Howard, C., Muno, M., Baganoff, F. K., Park, S., Feigelson, E., 
    Garmire, G., Brandt, N., 2004, in 'The Dense Interstellar Medium in Galaxies', 
    Proc. of the 4th Cologne-Bonn-Zermatt Symposium, Zermatt, Switzerland, 
    22-26 September 2003. Edited by S.Pfalzner, C. Kramer, C. Staubmeier, and A. Heithausen. 
    Springer proceedings in physics, Vol. 91. Berlin, Heidelberg: Springer, 2004, p.281}
\rf{Morris, M., 1996, Nature, 383, 389}
\rf{Morris, M., Serabyn, E., 1996, ARA\&A 34, 645}
\rf{Nord, M.E., Lazio, T.J.W., Kassim, N.E., Hyman, S.D., 
    LaRosa, T.N., Brogan, C.L., Duric, N., 2004, 	AJ 128, 1646}
\rf{Ott et al. 1999 ApJ 523, 248}
\rf{Paumard, T., Maillard, J.-P., Morris, M., 2004, A\&A 426, 81}
\rf{Pollack, L. K., Taylor, G. B., Zavala, R. T., 2003, ApJ 589, 733}
\rf{Pushkarev, A.B., Gabuzda, D. C., Vetukhnovskaya, Yu.N., Yakimov, V.E., 2005, MNRAS 356, 859}
\rf{Rusk, R., 1988, in 'The Impact of VLBI on Astrophysics and Geophysics',
   Proceedings of the 129th IAU Symposium, Cambridge, MA, May 10-15, 1987.
   Edited by Mark Jonathan Reid and James M. Moran.,
   Dordrecht, Kluwer Academic Publishers, p.161}
\rf{Sch\"odel, R., Genzel, R., Ott, et al. 2003, ApJ, 596, 1015}
\rf{Sch\"odel, R., Ott, T., Genzel, R. et al. 2002, Nature, 419, 694}
\rf{Shakura, N.I.,\& Sunyaev, R.A., 1973, A\&A, 24, 337}
\rf{Torres, D.F., Capozziello, S., \& Lambiase, G., 2000, Phys.Rev. D, 62, 104012}
\rf{Vollmer \& Duschl 2000, New Astron., 4, 581}
\rf{Yuan, Y.-F., Narayan, R. \& Rees, M.J.,   2004, ApJ 606, 1112}
\rf{Yuan, F., Quataert, E. \& Narayan, R. 2004, ApJ,  606, 894}
\rf{Yuan, F., Markoff, S. \& Falcke, H. 2002, A\&A, 854, 854}
\rf{Yusef-Zadeh, F., Roberts, D.A., Biretta, J., 1998, ApJ 499, L159}
\rf{Yusef-Zadeh, F., et al., 2006, ApJ in press, astro-ph/0510787}

\newpage


\begin{figure}
\centering
\includegraphics[angle=-00,width=12cm]{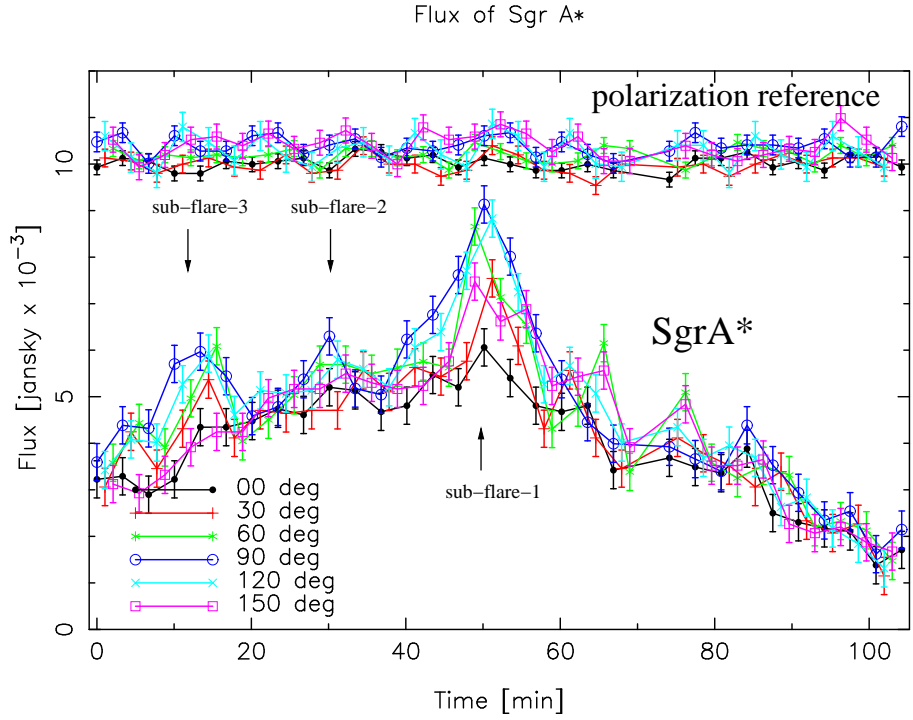}
\caption{\small
\label{Fig:1}
Flux density at different PAs
of the Wollaston prism  for the polarization reference star
$\sim$0.38'' north of SgrA* (Fig.\ref{Fig:2}) and
the NIR counterpart of SgrA*
on the nights between 29 and 30 July 2005. 
Six polarization angles 
0$^o$, 30$^o$, 60$^o$,  90$^o$, 120$^o$, and 150$^o$ (East of North) 
were used.
For display  purposes the flux density values of the reference 
star have been shifted upwards by 5~mJy.
}
\end{figure}

\begin{figure}
\centering
\includegraphics[angle=-00,width=12cm]{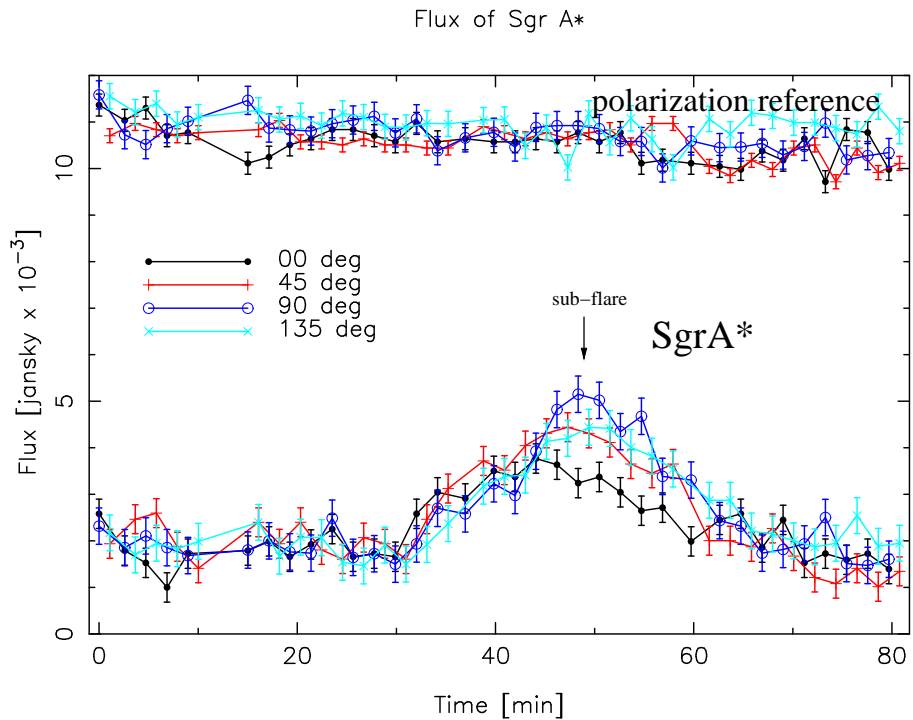}
\caption{\small
\label{Fig:2}
Flux density at different PAs
of the Wollaston prism for a polarization reference star
near SgrA* and
the NIR counterpart of SgrA* taken
on the nights between 12 and 13 June 2004.
Four polarization angles 
0$^o$, 45$^o$, 90$^o$, and 135$^o$ (East of North) 
were used.
For display  purposes the flux density values of the reference 
star have been shifted upwards by 5~mJy.
}
\end{figure}

\FloatBarrier
\newpage

\begin{figure}
\centering
\includegraphics[angle=-00,width=15cm]{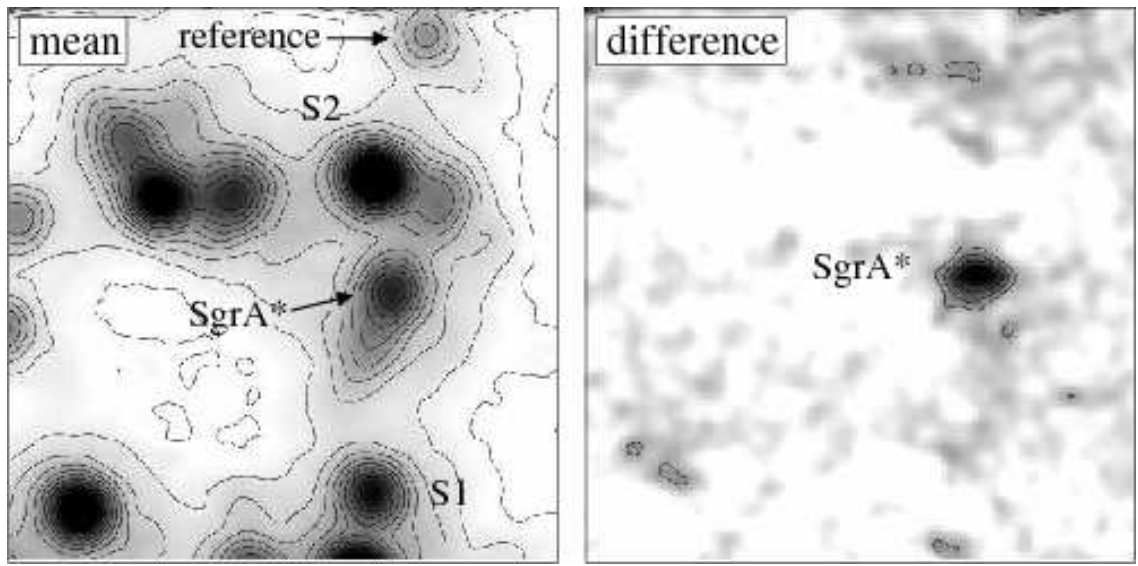}
\caption{\small 
\label{Fig:3}
The mean (left) and difference (right) of our K-band NACO images 
at 0$^o$ and 90$^o$ polarization angle 
taken simultaneously through the Wollaston prism 
during the peak of the flare at $t=$50 after the start of the
measurements in July 2005 (see Fig.\ref{Fig:1}).
}
\end{figure}

\FloatBarrier
\newpage

\begin{figure}
\centering
\includegraphics[angle=-00,width=11cm]{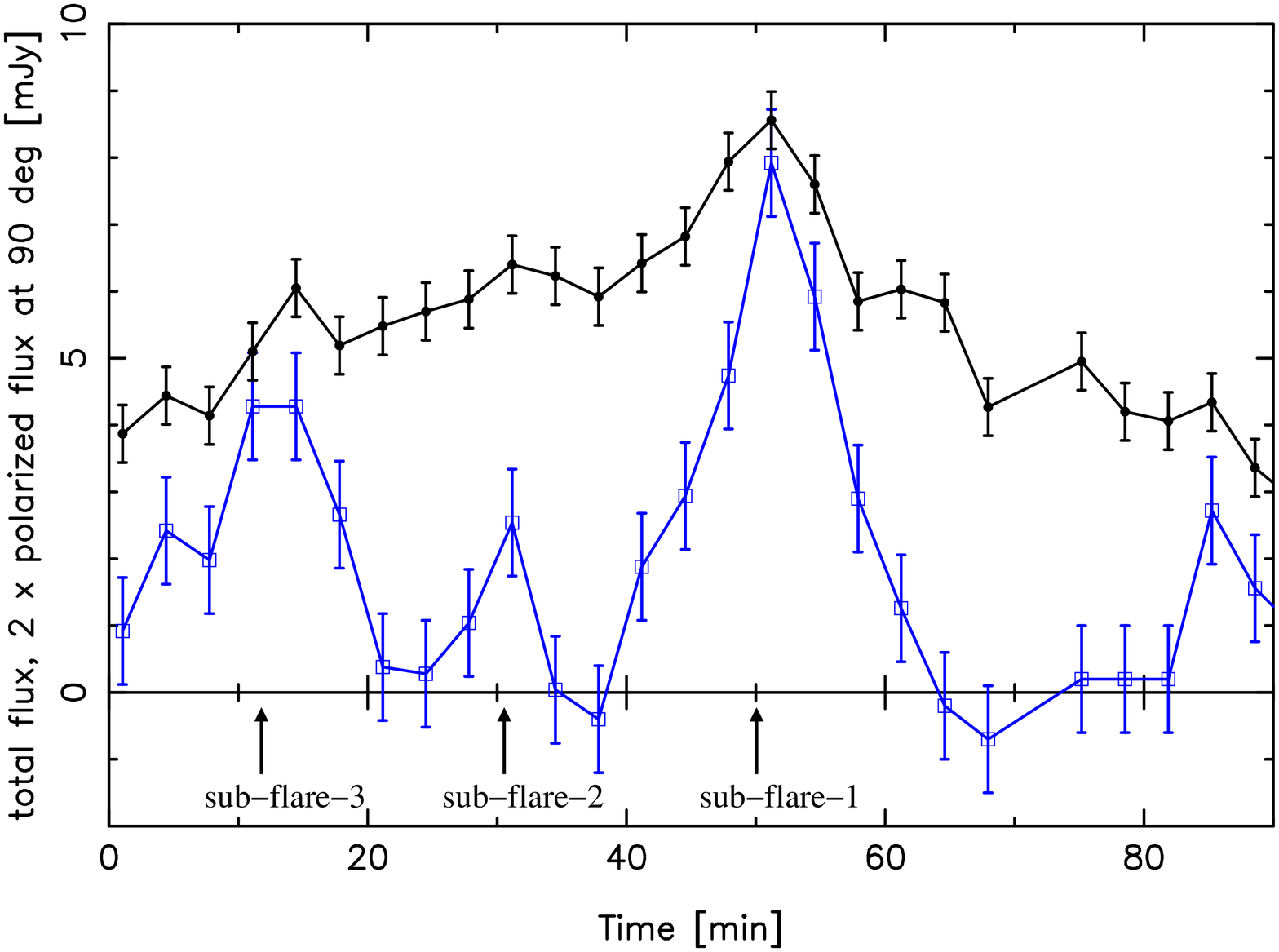}
\includegraphics[angle=-00,width=11cm]{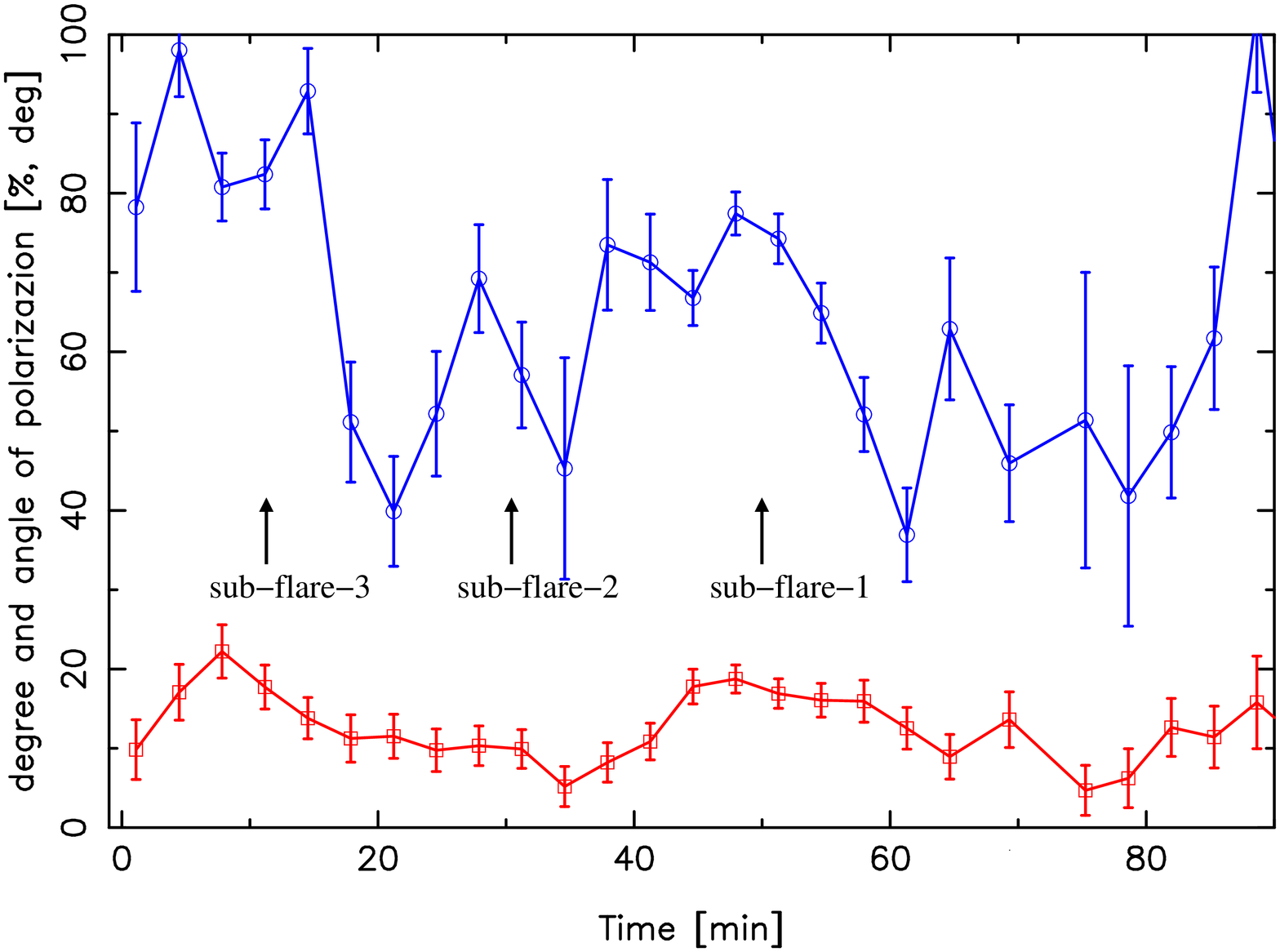}
\caption{\small
\label{Fig:4}
Polarized emission from the NIR counterpart SgrA* in July 2005.
{\bf Top:} The total de-reddened flux density light curve (black) and
the de-reddened flux density at polarization angle 90$^o$ 
(East over North; blue) corrected for the flux density measured 
at a PA of 0$^o$ at which the sub-flares cannot be seen.
{\bf Bottom:} 
The position angle of the E-vector (top graph, blue)
and the degree of polarization (bottom graph, red). 
The left axis is used to label both quantities in different units
as given in brackets.
The arrows are separated by 20 minutes (see Fig.\ref{Fig:6})
and mark the position of the three sub-flares. 
Another weak peak in polarized flux that is in agreement with the observed 
quasi-periodicity can be seen at $t=$85.
}
\end{figure}

\FloatBarrier
\newpage

\begin{figure}
\centering
\includegraphics[angle=-00,width=11cm]{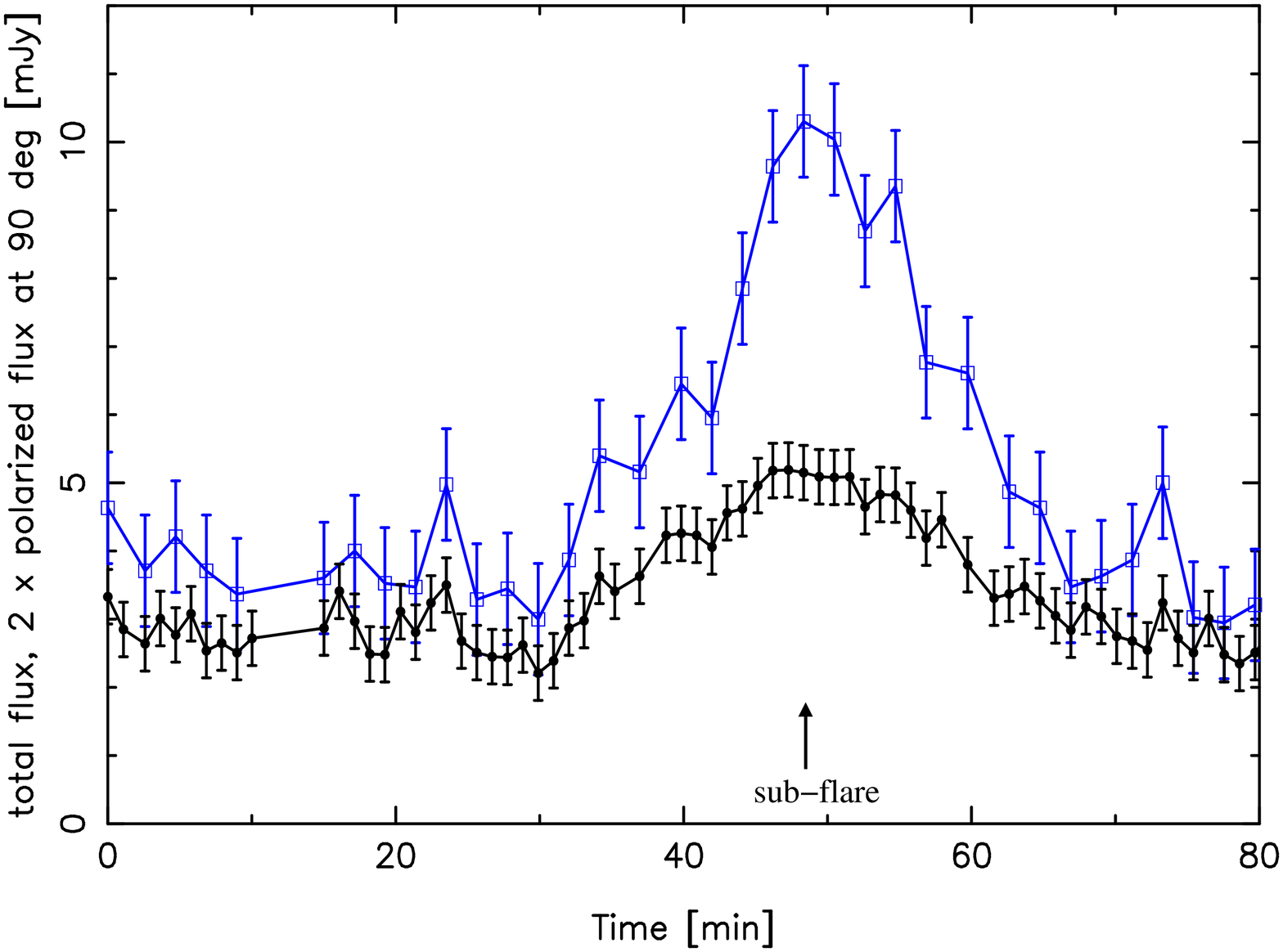}
\includegraphics[angle=-00,width=11cm]{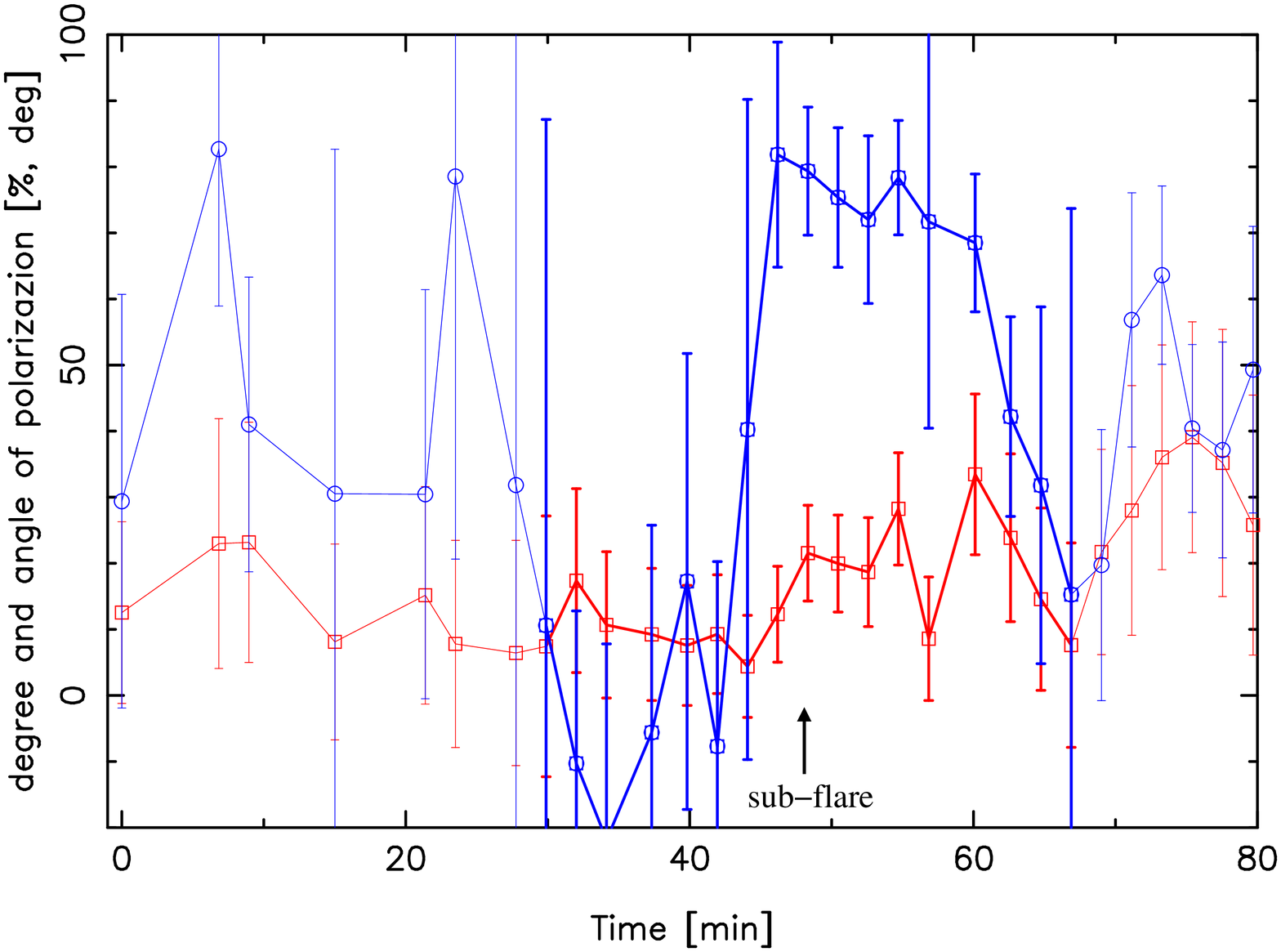}
\caption{\small
\label{Fig:5}
Polarized emission from the NIR counterpart SgrA* in June 2004.
{\bf Top:} The total de-reddened flux density light curve (black) and
the de-reddened flux density at polarization angle 90$^o$ 
(East over North; blue) corrected for the flux density measured 
at a PA of 0$^o$ at which the sub-flares cannot be seen.
{\bf Bottom:} 
The position angle of the E-vector (top graph, blue)
and the degree of polarization (bottom graph, red). 
The left axis is used to label both quantities in different units
as given in brackets.
The arrow marks the position of the polarized sub-flare. 
The region over which the flare emission is strong have been
highlighted by bold face lines.
Over the remaining sections of the plot the flare emission is weak and the
error bars are correspondingly large. For completeness, however, we have included 
these data points as well.
}
\end{figure}

\FloatBarrier
\newpage

\begin{figure}
\centering
\includegraphics[width=14cm]{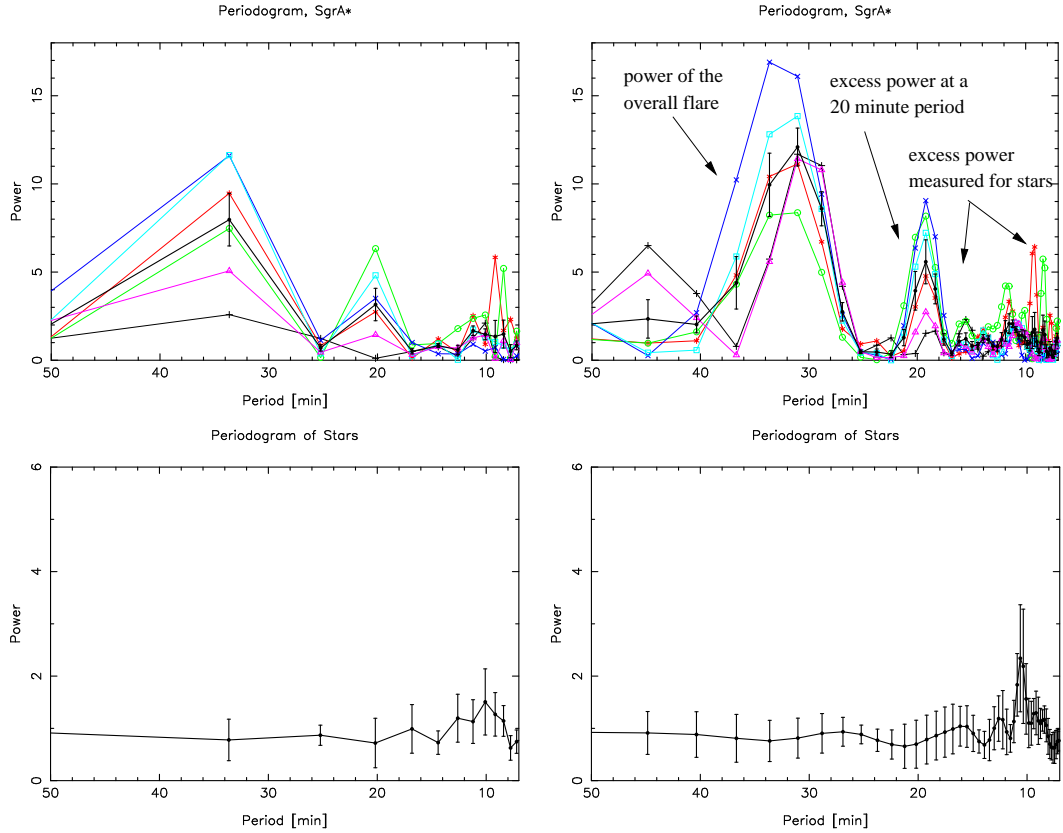}
\caption{\small 
\label{Fig:6}
{\bf Top left:} Periodogram of the July 2005 polarized 
flux density data from SgrA* 
at individual polarization angles 
(see color coding information in Fig.\ref{Fig:1})
 and for the mean (black).
{\bf Bottom left:} 
Mean and standard deviation calculated from periodograms of 7 stars 
within 1.5 arcsecond radius of SgrA*.
{\bf Top and Bottom right:} The same as top and bottom right but
oversampled in order to allow a better estimate of the peak widths.
Note that both the mean and PA 0$^o$ curves are printed in black.
They can, however, be clearly distinguished from each other since only the data 
points of mean curve have error bars associated with them.
}
\end{figure}

\FloatBarrier
\newpage

\begin{figure}
\centering
\includegraphics[width=13cm,angle=-90]{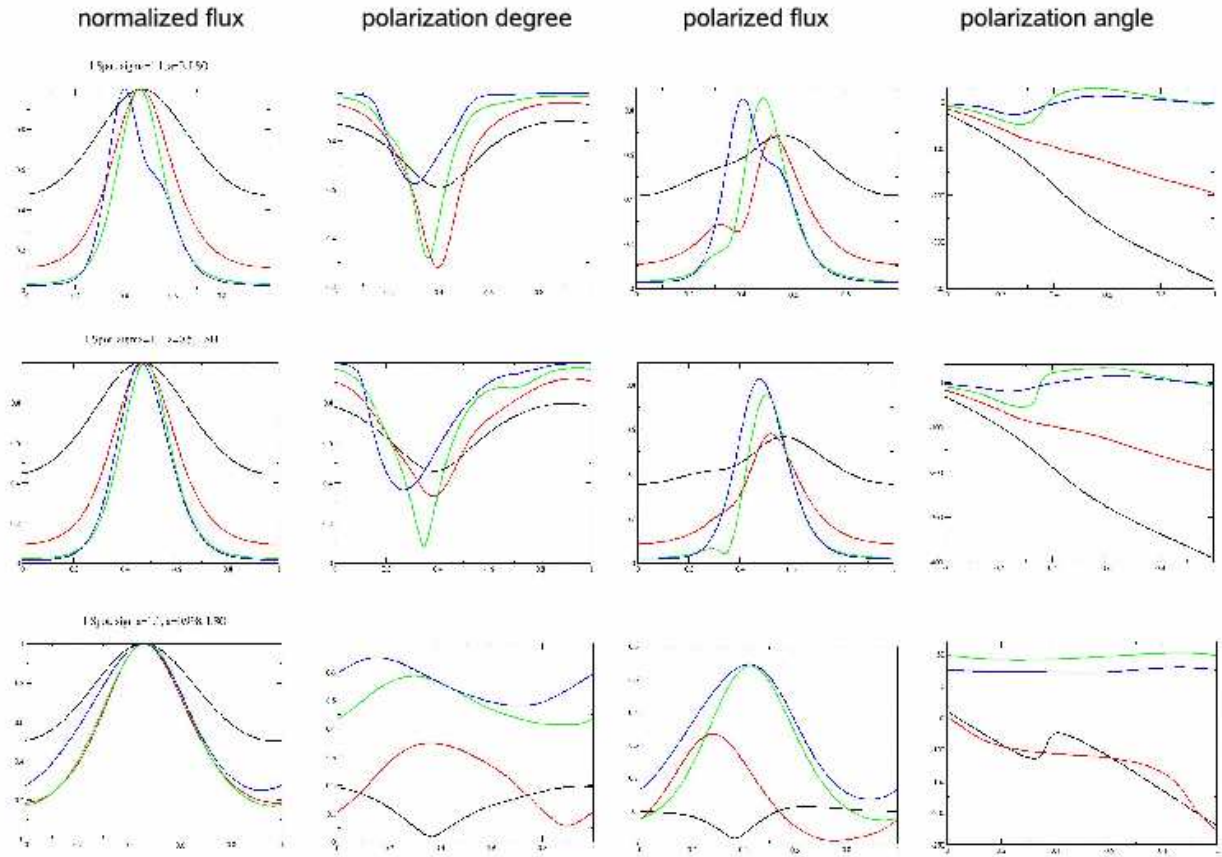}
\caption{\small 
\label{Fig:7}
Curves of peak normalized flux density, degree of polarization, polarized 
flux density and PA of the E-vector as calculated for 
orbiting spot models 
based on the code by Dovciak, Karas, \& Yaqoob (2004).
The data are shown for one orbital period.
The inclinations are color coded as 
10$^o$  black,
30$^o$  violet,
50$^o$  green,
70$^o$  blue.
}
\end{figure}

\FloatBarrier
\newpage

\begin{figure}
\centering
\includegraphics[angle=-00,width=15cm]{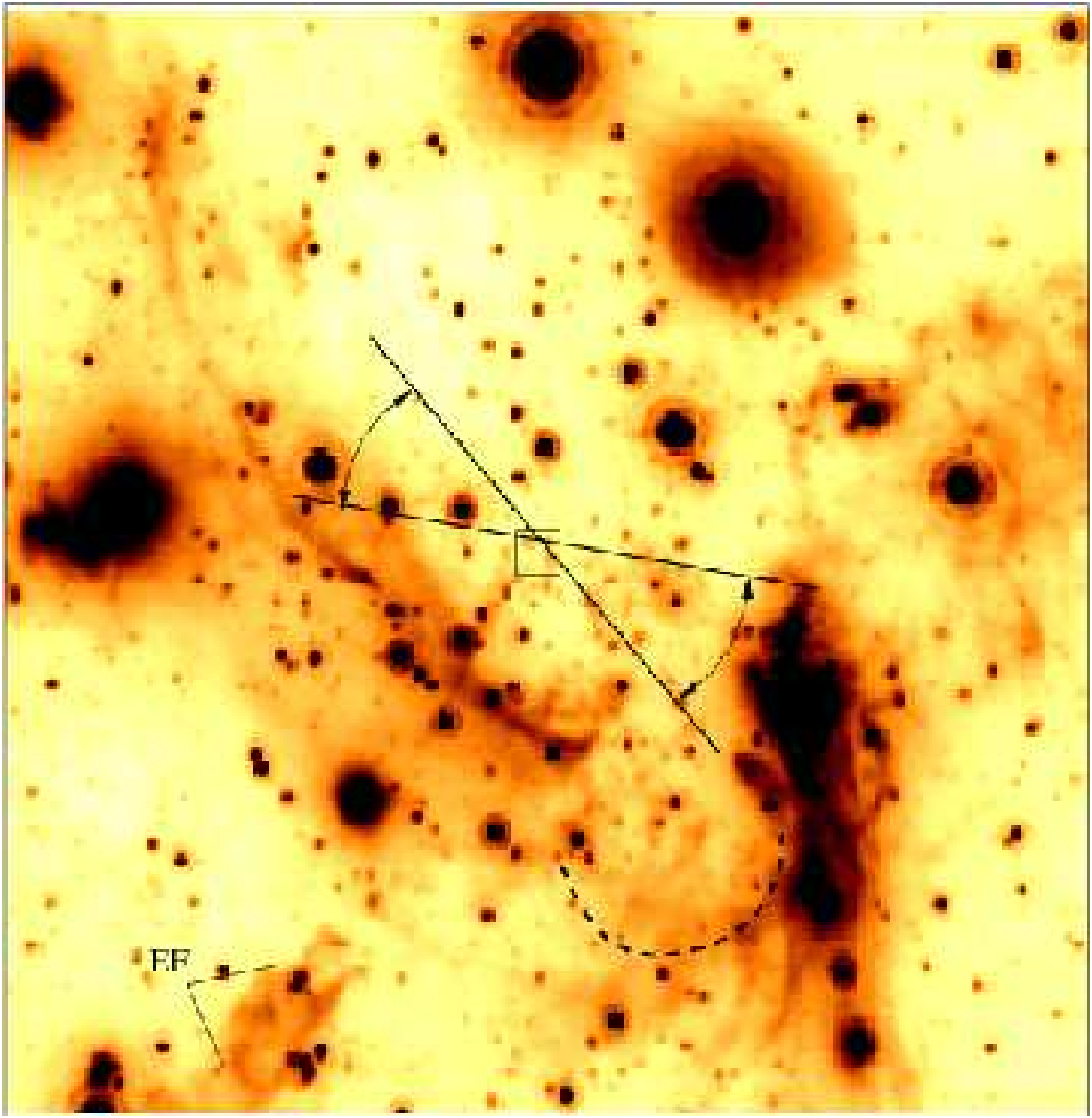}
\caption{\small 
\label{Fig:8}
An image of the central 0.5$\times$0.5 square parsec 
(1 arcsec $\sim$ 0.039 parsecs) of the Galactic Center at a 
wavelength of 3.8$\mu$m using the NACO adaptive optics system at the VLT.
In addition to the stars the dust emission from the min-spiral is seen.
The rectangular box indicates the image section shown in Fig.\ref{Fig:3}. 
The two lines centered on the position of SgrA* cover the 
range over which the polarization angle 
observed in July 2005 and June 2004 varies on the sky.
The approximate location and shape of the mini-cavity is shown 
as a dashed line.
The source EF is described in the text and marked in Fig.\ref{Fig:9}.
}
\end{figure}

\FloatBarrier
\newpage

\begin{figure}
\centering
\includegraphics[angle=-00,width=12cm]{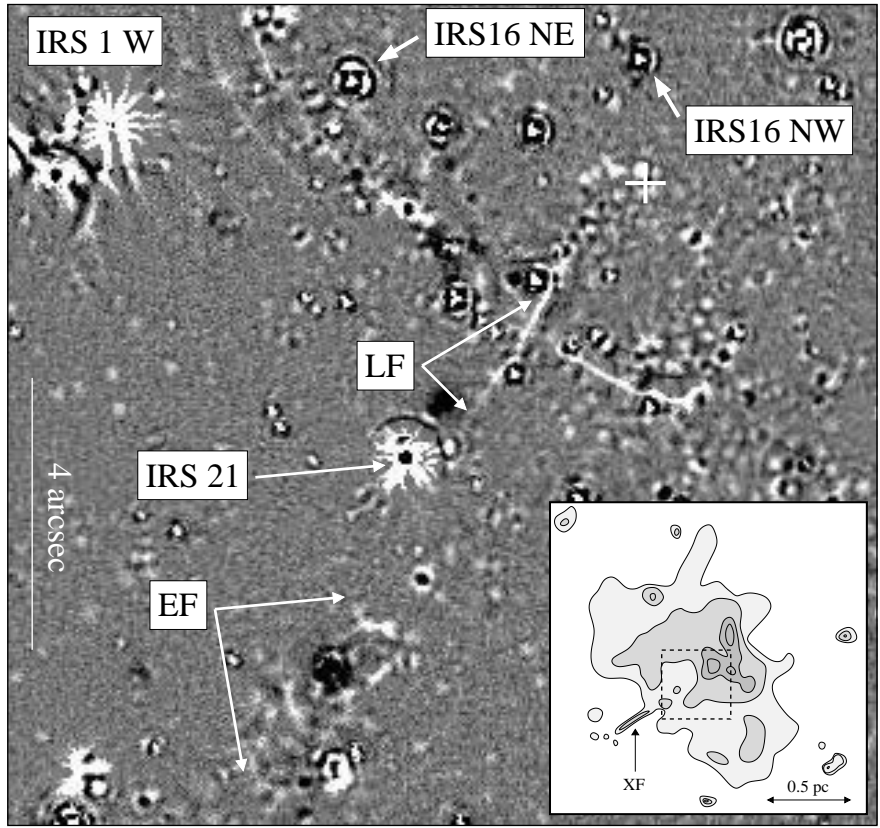}
\caption{\small 
\label{Fig:9}
High pass filtered L'-band image (centered at 3.8$\mu$m wavelength)
in which the majority of the extended and 
stellar flux has been subtracted using the {\it StarFinder} code
(Diolaiti et al. 2000), i.e the shown image is a {\it StarFinder} residual image.
Some residuals are left at the positions of 
bright stars, some of which have been labeled. 
The position of SgrA* is marked by the red cross 
in the north-western quadrant of the map.
In addition to a few sharp edged structures associated with the
northern arm dust emission of the mini-spiral one can see an linear 
feature (LF) and an extended structure (EF) both pointing
approximately into the direction of SgrA* (see also Fig.\ref{Fig:8}).
The inset shown in the lower right corner is a sketch of the grey scale
image shown in Morris et al. (2004). The X-ray feature discussed in the text is labeled as XF.
The NIR filament LF and the X-ray filament XF are not colinear.
The dashed rectangular line markes the position and extent of the 
high pass filtered L'-band image shown here.
}
\end{figure}

\FloatBarrier
\newpage

\begin{figure}
\centering
\includegraphics[angle=-00,width=10.7cm]{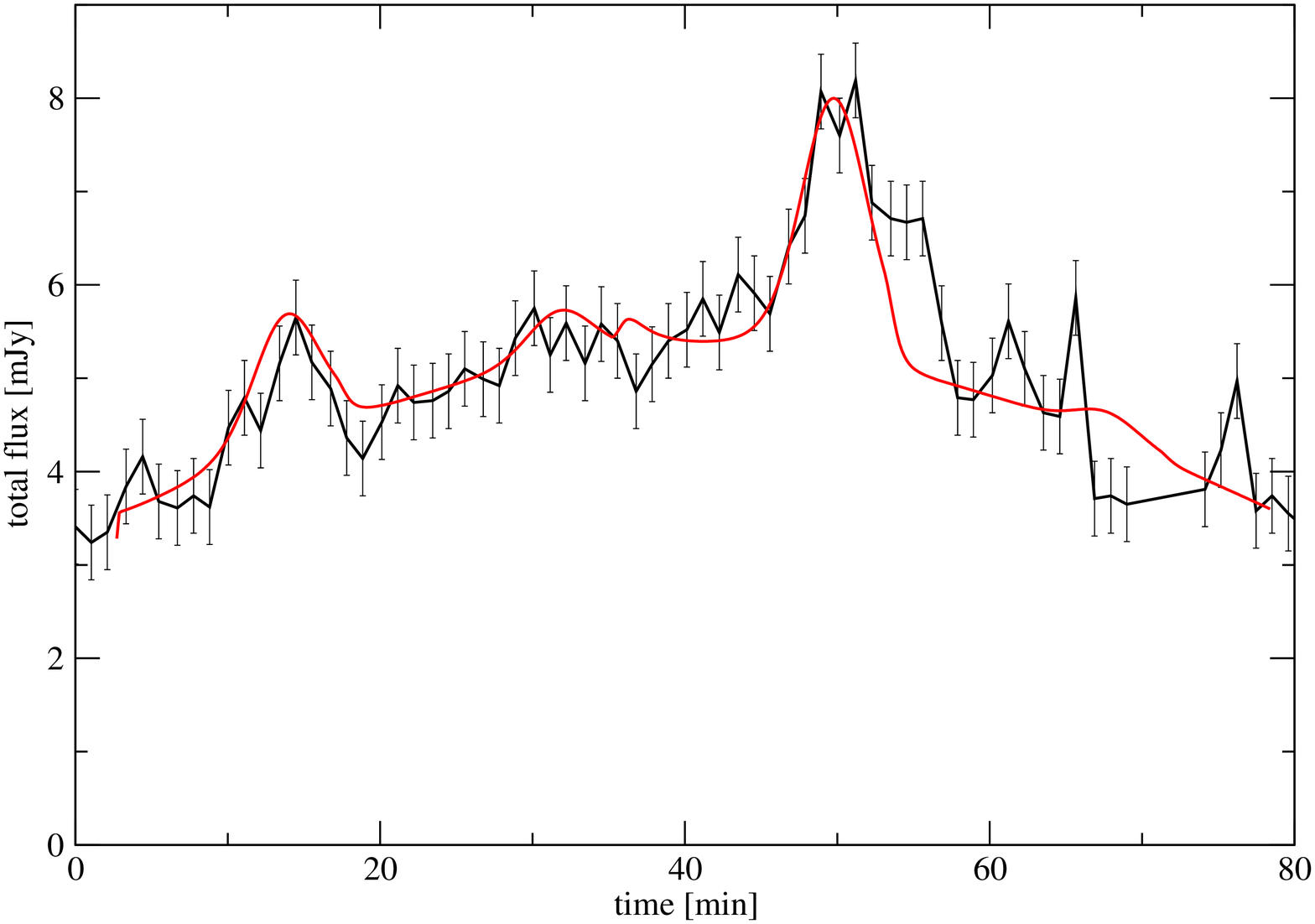}
\includegraphics[angle=-00,width=11cm]{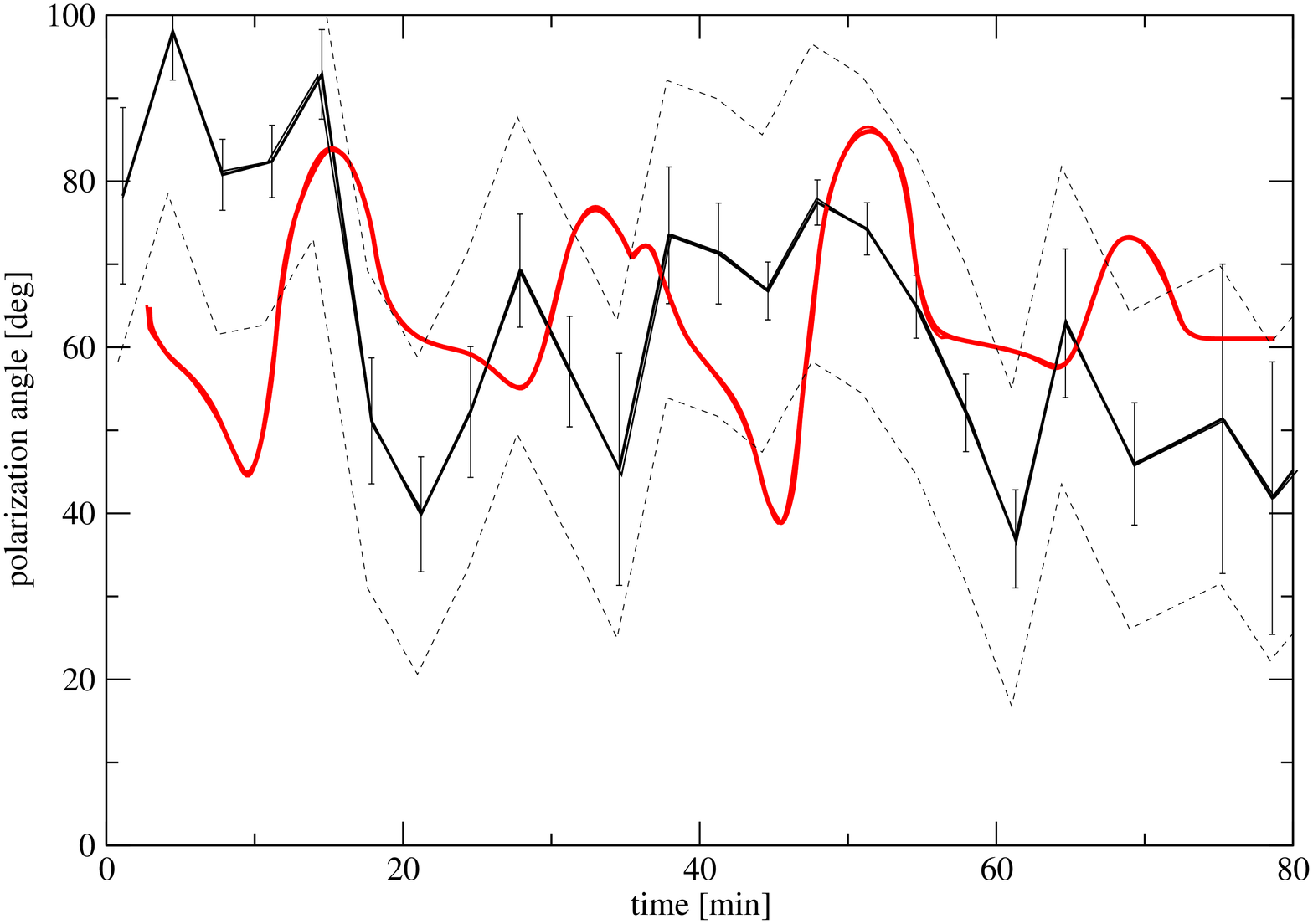}
\caption{\small 
\label{Fig:10}
A comparison between an orbiting spot model (red lines)
for July 2005 and the
measured total flux density (top) and polarization angle (bottom).
As a demonstration of the consistency between the model and the data 
we have chosen a solution that gives a satisfactory representation 
of the total flux density.
For the chosen inclination of 55$^o$ the position angle of the E-vector
resulting from the model and interstellar polarization does not wrap
and agrees with the measurements to within $\pm$3 times the mean 1$\sigma$
uncertainties of the measurements - as indicated by the two thin dashed lines.
This solution has to be compared to senarii with different inclinations
as shown in the model results in Fig.\ref{Fig:7}.
}
\end{figure}

\begin{figure}
\centering
\includegraphics[width=14cm]{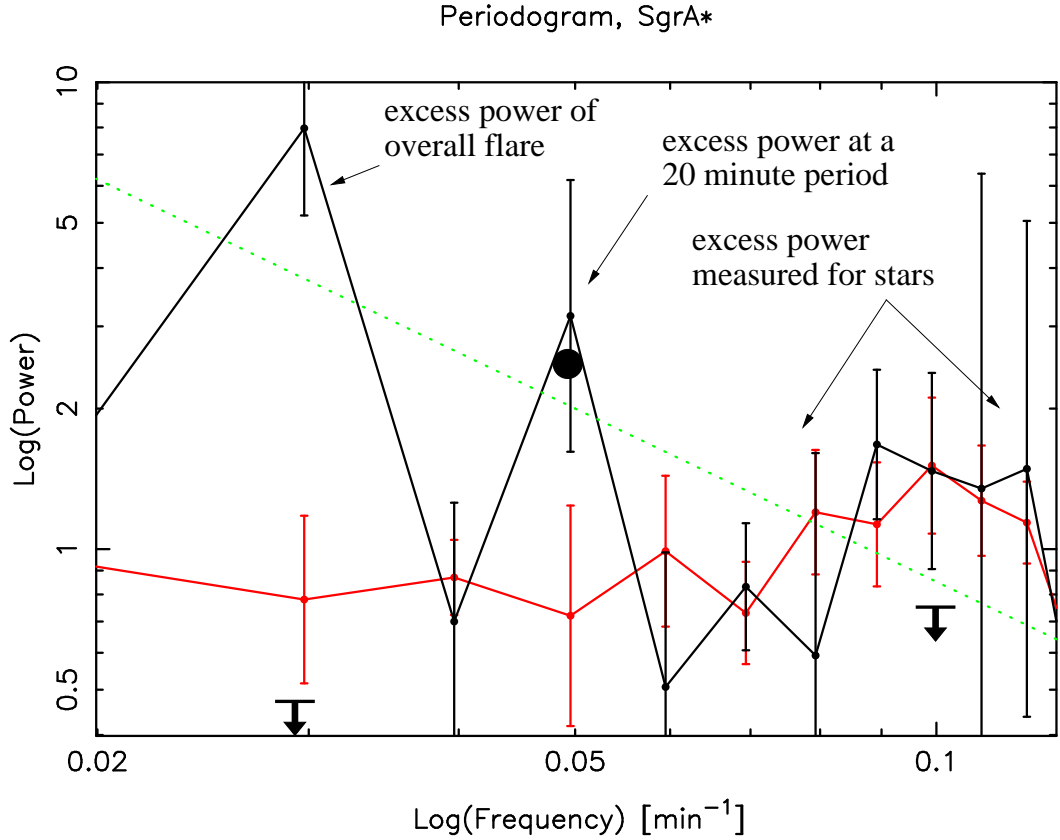}
\caption{\small 
\label{Fig:11}
{\bf Top right:} 
Double logarithmic periodogram of the July 2005 polarized flux density data from SgrA* 
at individual polarization angles.
The data for SgrA* is shown in black and the 
mean and standard deviation calculated from periodograms of 7 stars 
within 1.5 arcsecond radius of SgrA* are shown in red.
The green dashed line corresponds to a fit of the data by a power spectrum 
of the form $P(f) \sim f^{-\beta}$ with an exponent
$\beta = 1.2\pm0.7 (1\sigma)$.
Hence our July 2005 NIR data are consistent with a 
flat noise distribution ($\beta = 0$) and
do not provide an indication for a significant 
low frequency red-noise excess. 
The value and upper limits printed in bold face are explained in the text.
}
\end{figure}

\FloatBarrier
\newpage

\end{document}